\documentclass[preprint,superscriptaddress,showpacs,preprintnumbers,amsmath,amssymb,aps,prl]{revtex4}
\usepackage{graphicx}
\usepackage{dcolumn}
\usepackage{bm}
\begin{document}
\title{Entropy of seismic electric signals: Analysis in natural time under time-reversal}
\author{P. A. Varotsos}
\email{pvaro@otenet.gr}
\affiliation{Solid State Section, Physics Department, University of Athens, Panepistimiopolis, Zografos 157 84,
Athens, Greece}
\affiliation{Solid Earth Physics Institute, Physics Department, University of Athens, Panepistimiopolis, Zografos 157 84, Athens, Greece}
\author{N. V. Sarlis}
\affiliation{Solid State Section, Physics Department, University of Athens, Panepistimiopolis, Zografos 157 84,
Athens, Greece}
\author{E. S. Skordas}
\affiliation{Solid State Section, Physics Department, University of Athens, Panepistimiopolis, Zografos 157 84,
Athens, Greece}
\affiliation{Solid Earth Physics Institute, Physics Department, University of Athens, Panepistimiopolis, Zografos 157 84, Athens, Greece}
\author{H. K. Tanaka}
\affiliation{Earthquake Prediction Research Center, Tokai University 3-20-1, Shimizu-Orido, Shizuoka 424-8610, Japan}
\begin{abstract}
Electric signals have been recently recorded at the Earth's surface with amplitudes appreciably larger than those hitherto reported. 
Their entropy in natural time is smaller than that, $S_u$, of a ``uniform'' distribution. The same holds for their entropy upon time-reversal.  This behavior, as supported by numerical simulations in fBm time series and in an on-off intermittency model, stems from infinitely ranged long range temporal correlations and hence these signals are probably Seismic Electric Signals (critical dynamics). The entropy fluctuations are found to increase upon approaching bursting, which reminds the behavior identifying sudden cardiac death individuals when analysing their electrocardiograms.
\end{abstract}
\maketitle

The time series analysis of various phenomena in complex systems (and especially those associated with impending catastrophic events, e.g.\cite{YUL00,YUL01}) in the framework of the newly defined time-domain\cite{NAT01,NAT02}, termed natural time, reveals interesting features. Examples are electrocardiograms\cite{NAT04,NAT05},  seismicity\cite{NAT01} and Seismic Electric Signals (SES) activities\cite{NAT01,NAT02,NAT03,NAT03B,NAT05B,newbook}. This new time domain  is optimal for enhancing the signals in time-frequency space when employing the Wigner function and measuring its localization property\cite{ABE05}.

In a time series comprising $N$ pulses, the natural time $\chi_{k} = k/N$ serves as the index\cite{NAT01,NAT02} for the occurrence of the $k$-th event. It is, therefore, smaller than, or equal to, unity. In natural time analysis, the time evolution of the pair of the two quantities ($\chi_k, Q_k$)
 is considered, where $Q_k$ denotes the  duration of the $k$-th pulse. 
The entropy $S$ in the natural time-domain is defined\cite{NAT03B} as the derivative of the function $\langle \chi^q \rangle - \langle \chi \rangle ^q$ in respect to $q$, for $q=1$, which gives \cite{NAT01,NAT03B}:
\begin{equation}
S \equiv  \langle \chi \ln \chi \rangle - \langle \chi \rangle \ln  \langle \chi \rangle
\end{equation}
where $\langle f( \chi) \rangle = \sum_{k=1}^N p_k f(\chi_k )$ 
and $p_k=Q_{k}/\sum_{n=1}^{N}Q_{n}$. It is a dynamic entropy depending on the sequential order of pulses\cite{NAT04,NAT05} and exhibits\cite{NAT05B} concavity, positivity and Lesche\cite{LES82,LES04} stability.
When the system enters the critical stage (infinitely ranged long range temporal correlations\cite{NAT01,NAT02}), the $S$-value is smaller\cite{NAT03B,newbook} than the value $S_u (=1/2\ln 2-1/4\approx 0.0966$) of a ``uniform'' distribution (defined in Refs. \cite{NAT01,NAT02,NAT03B,NAT05}, e.g.   when when all $p_k$ are equal), i.e.,
\begin{equation}
S < S_u
\end{equation}
The value of the entropy obtained upon considering the time reversal ${\cal T}$, i.e., ${\cal T} p_k=p_{N-k+1}$, is labelled by $S_-$. An important point emerged from the data analysis in Ref.\cite{NAT05B} is the following: Although the study of the $S$ values enables the distinction between SES activities and noises produced by nearby operating artificial (man made) electromagnetic sources (AN), i.e., $S<S_u$ for the SES activities and $S \gtrapprox S_u$ for AN, this does not hold for the $S_-$ values. This is so, because for the SES activities we found that the $S_-$ values are smaller than (or equal to) $S_u$, while for AN the $S_-$ values are either smaller or larger than $S_u$. Here, we provide more recent data on the SES activities, which strengthen the conclusion that both $S$ and $S_-$ are smaller than $S_u$. In other words, the following key point seems to hold: In signals that exhibit critical dynamics (i.e., SES activities) upon time-reversal their entropy values (though become {\em different} than those in forward time) still remain smaller than $S_u$. Why? The answer to this question is a challenge, because, if it is generally so, among similar looking signals we can distinguish those that exhibit critical dynamics. Since the latter signals are expected to exhibit infinitely ranged long range correlations, this might be the origin of the aforementioned behavior. To investigate this point, numerical simulations are presented here  for fractional Brownian motion (fBm) time series as well as for an on-off intermittency model (for the selection of the latter see Ref.\cite{EPAPS}). The simple case of an fBm was selected in view of the following suggestion forwarded in Ref.\cite{NAT03} concerning the Hurst exponent $H$.
If we assume that, in general, $H$ is actually a measure of the intensity of long range dependence, we may understand why in SES activities, when analyzed in the natural time domain, we find $H$ values close to unity, while in AN (where the long range correlations are weaker\cite{NAT03}) the $H$-values are markedly smaller (e.g., around 0.7 to 0.8).

We first present the recent experimental results.  Figure 1 depicts five electric signals, labelled $M_1$,$M_2$, $M_3$, $M_4$ and $V_1$, that have been recorded on March 21, 23 and April 7, 2005. Note that the first four ($M_1$ to $M_4$) have amplitudes  that not only are one order of magnitude larger than the fifth one ($V_1$, see Fig.1c), but also significantly exceed those hitherto reported\cite{VAR03}. The way they are read in natural time can be seen in Fig.2. The analysis of these signals leads to the $S$ and $S_-$ values given in Table \ref{tab1}, an inspection of which reveals that they are actually smaller than $S_u$. Hence, on the basis of the aforementioned criterion, these signals {\em cannot} be classified as AN, but they are likely to be SES activities. Note that, although in general $S$ is different than $S_-$, no definite conclusion can be drawn on the sign of $S-S_-$ (see also Table I of Ref.\cite{NAT05B}).

We now present our results on fBm. We first clarify that Weron et al.\cite{WER05} recently studied an algorithm distinguishing between the origins (i.e., the memory and the tails of the process) of the self-similarity of a given time series on the base of the computer test suggested in Ref.\cite{MER03}. By applying it to the SES activities, they found the fBm as the appropriate type of modelling process.  The fBm, which is $H$-self similar with stationary increments and it is the only Gaussian process with such properties
 for $0 < H < 1$\cite{SAM94}, can be simulated\cite{MAN69,SZU01}, see 
 also pp.321 -323 of Ref.\cite{MAN01},  by 
randomizing a construction  due to Weierstrass, i.e.,  using the 
Weierstrass-Mandelbrot function\cite{inter}:
\begin{equation}
\label{weiman}
w(t)=\sum_{l} c_l \frac{\sin (b^lt+d_l)}{b^{lH}},
\end{equation}
where $b>1$, $c_l$ normally distributed with mean 0 and standard deviation 1, and $d_l$
are uniformly distributed in the interval $[0,2\pi]$
(cf. when using the increments of Eq.(\ref{weiman}) one can 
also produce fractional Gaussian noise of a given $H$).
 By using Eq. (\ref{weiman}),
fBm for various values of $H$ were produced,
 the one-signed segments of which 
were analyzed in the natural time domain (an example is given in Ref.\cite{EPAPS}). 
The Monte-Carlo simulation
results for each segment  include not 
only the values of the entropies $S$ and $S_-$, 
but  also the exponent $\alpha_{DFA}$ of the Detrended Fluctuation Analysis (DFA)\cite{p18,p19}.  For segments of small number 
of points $N$ (cf. only segments with $N>40$ were considered), the value of $\alpha_{DFA}$ may vary significanly from that 
expected for a given value of $H$; DFA was 
preferred, because it is one of the few well-defined and robust estimators  of the scaling properties 
for such segments(e.g. \cite{NAT03}, see also pp. 300-301 of Ref.\cite{newbook}).
The results are shown in Fig.3, in which we plot the $S$ and $S_-$ values versus 
 $\alpha_{DFA}$. Since the analysis of the SES activities
  in natural time result\cite{NAT03,NAT03B} in DFA exponents 
    $\alpha_{DFA}$ around unity, hereafter we are 
  solely focused in Fig.3 in the range $0.8<\alpha_{DFA}<1.2$.
   An inspection of this figure reveals the following 
   three conclusions: First, despite 
   the large standard deviation, we may say that 
   both $S$ and $S_-$ are smaller 
   than $S_u (\approx 0.0966)$ 
   when $\alpha_{DFA} \approx 1$. Second, $S$ and $S_-$
    are more or less comparable. Third, 
   comparing the computed $S$ and $S_-$ values
    ($\approx 0.08$ for $\alpha_{DFA}\approx 1$) with 
   those resulting from the analysis of 
   the SES activities (see Table \ref{tab1}, see also Table I of Ref.\cite{NAT05B}),
    we find a reasonable agreement. 
    Note that these computations do {\em not} result 
    in a definite sign for $S - S_-$ in a similar 
    fashion with the experimental results.  

In what remains, we  present our results on a simple on-off intermittency model. We clarify that on-off intermittency is a phase-space mechanism that allows dynamical systems to undergo bursting (bursting is a phenomenon in which episodes of high activity are alternated with periods of inactivity likewise in Fig.2e). This mechanism is different from the well known Pomeau-Manneville scenario for the behavior of a system in the proximity of a saddle-node bifurcation\cite{POM80}. Here, we use the simple model of the driven logistic map
\begin{equation}
\label{eq11}
X_{t+1} = A(Y_t) X_t [1-X_t]
\end{equation}
where we assume that the quantity $A(Y_t)$ is {\em monotonic} function of $Y_t$ and that $0 \leq A \leq 4$ (cf. $A$ is further specified below). The system has the invariant manifold $X=0$ and the level of its activity is measured by $X_t$\cite{TON02}. In order to have the on-off mechanism in action, we specialize to the case of a noise-driven logistic map, with
\begin{equation}
\label{eq12}
A(Y_t) = A_0 + \alpha Y_t
\end{equation}
where $Y_t$ is $\delta$-correlated noise which is uniformly-distributed in the interval [0,1] and $A_0$ and $\alpha$ are parameters. In order to have $0 \leq A \leq 4$, we assume\cite{TON02} $A_0 \geq 0, \alpha \geq 0$ and $A_0+ \alpha \leq 4$. The relevant parameter plane for the noise-driven system of Eqs.(\ref{eq11}) and (\ref{eq12}) (as well as the parameter range for which the fixed point $X=0$ is stable) can be found in Fig. 1 of Ref.\cite{TON02}, while the description of the intermittent dynamics is given in Refs.\cite{PLA93,HEA94,BAL99}. Bursting is observed in the temporal evolution of $X_t$ as the stability of the fixed point $X=0$ varies. Following Ref.\cite{HEA94}, for $A_0=0$ there is a {\em critical} value $\alpha_c > 1$, below which the system asymptotically tends to the fixed point $X=0$, without any sustained intermittent bursting. For this case, i.e., $A_0=0$, the value $\alpha_c =e\equiv 2.71828\ldots$ leads to on-off intermittency\cite{TON02}. In the intermittent system under discusssion,  both the signal amplitude and the power spectrum resulted\cite{TON02} in power-law distributions (with low frequencies predominating in the power-spectrum)

Several time-series have been produced for the above on-off intermittency model with the following procedure: The system was initiated at a time ($t_{in}=-200$)   with a uniformly distributed value $X_{t_{in}}$ in the region $[0,1]$, and then
the mapping of Eqs.(\ref{eq11}) and (\ref{eq12}) was followed until $N$ 
events will occur after $t=0$. The results for $X_t, t=1,2 \ldots N$ were analyzed in natural time domain and the values of $S$ and $S_-$ have been determined. This was repeated $10^3$ times (for a given number, $N$, of events) and the average values of $S$ and $S_-$ have been deduced. These values are plotted in Fig.4(a) versus $(\alpha - e) N^{1/2}$ (The factor $N^{1/2}$ stems from finite size scaling effects, since for large values of $N$, e.g., $N >$15000, a scaling -reminiscent of a 1st order phase transition- was observed, details on which   will be published elsewhere). This figure reveals that as the {\em critical} value is approached from below (i.e., $\alpha \rightarrow e_-$) both $S$ and $S_-$ are smaller than $S_u$. Note that Fig.4(a) also indicates that $S$ is probably larger than $S_-$, while in the fBm time series no definite sign for $S-S_-$ could be obtained.

Another interesting point resulted from the on-off intermittency model is depicted in Fig.4(b), where we plot the fluctuations $\delta S$ and $\delta S_-$ (i.e., the standard deviations of the entropies $S$ and $S_-$, respectively) versus $(\alpha - e) N^{1/2}$. It is clearly seen that these fluctuations are dramatically enhanced as the {\em critical} value is approached (i.e., $\alpha \rightarrow e$). This is strikingly reminiscent of our earlier results\cite{NAT04,NAT05} upon analyzing electrocardiograms (ECG) in natural time domain and studying the so called QT intervals. These results showed that the fluctuations of the entropy $\delta S(QT)$ are appreciably larger in sudden cardiac death (SD) individuals than those in truly healthy (H) humans (see Fig. 2 of Ref.\cite{NAT05}). We emphasize that the aforementioned points should not be misinterpreted as stating that the simple logistic map model treated here can capture the complex heart dynamics, but only can be seen in the following frame: Since sudden cardiac arrest (which may occur even if the electrocardiogram looks similar to that of H) may be considered as a dynamic phase transition\cite{NAT04, NAT05}, it is reasonable to expect that the entropy fluctuations significantly increase upon approaching the transition.

In summary, recently recorded electric signals (having the largest amplitudes recorded to date) exhibit the property that both $S$ and $S_-$ are smaller than $S_u$ and hence are likely to be SES activities (critical dynamics). This property seems to stem from their infinitely ranged long range correlations as supported by computational results in: (1) fBm time series and (2) a simple on-off intermittency model. The latter model also suggests that the fluctuations ($\delta S$ and $\delta S_-$) significantly increase upon approaching the transition, which is strikingly reminiscent of the increased $\delta S$-values found for the QT-intervals for the sudden cardiac death individuals.


\begin{figure}
\includegraphics{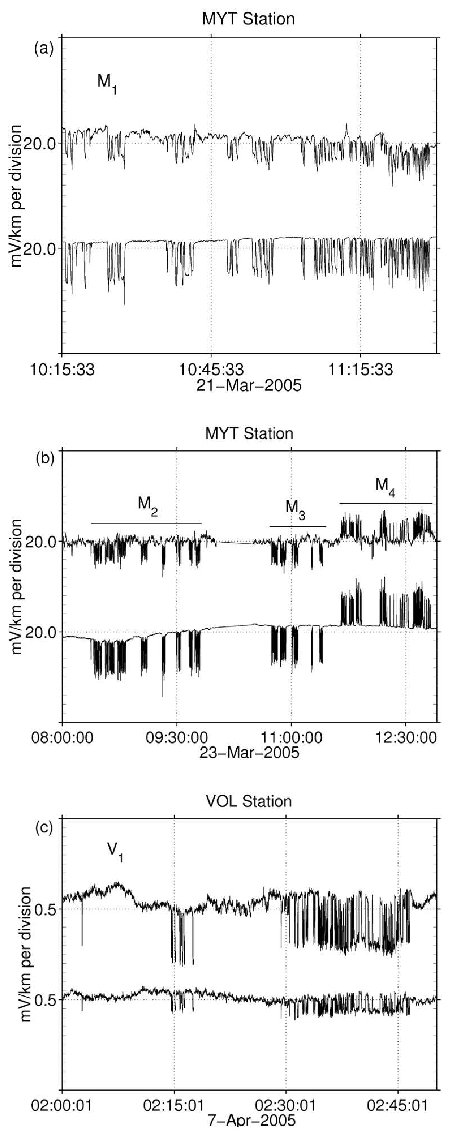}
\caption{\label{Fig1} Electric Signals recorded on March 21, 2005 (a), March 23, 2005 (b) and April 7, 2005 (c). The signals in (a) and (c) are labeled hereafter $M_1$ and $V_1$ respectively, while that in (b) consists of the three signals' activities labeled $M_2$, $M_3$, $M_4$ (sampling frequency $f_{exp}$=1Hz). The Universal Time (UT) is marked on the horizontal axis. Additional details for the two dipoles -records of which are shown here- as well as for the sites of the measuring stations are provided in Ref.\cite{EPAPS}.}
\end{figure} 

\begin{figure}
\includegraphics{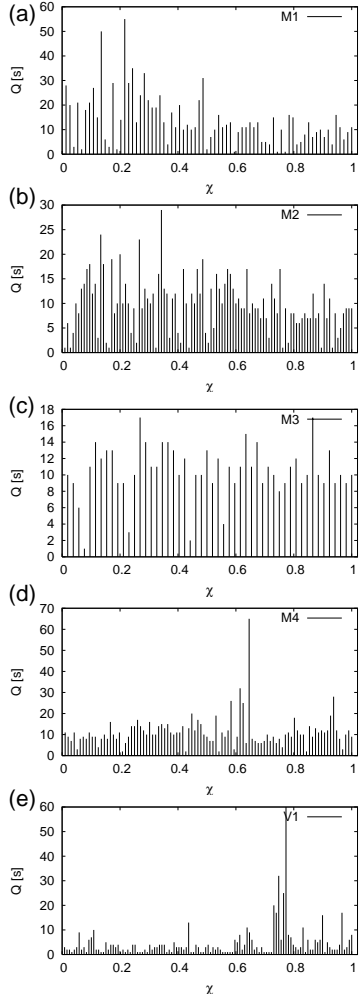}
\caption{\label{fig2} How the signals depicted in Fig.1 are read in natural time. (a), (b), (c), (d), (e), correspond to the signals' activities labeled $M_1$, $M_2$, $M_3$, $M_4$ and $V_1$, respectively.}
\end{figure}

\begin{figure}
\includegraphics{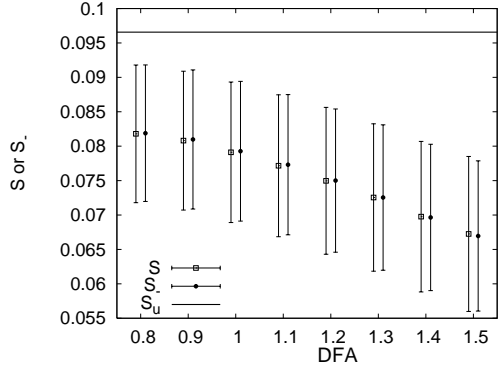}
\caption{\label{fig4} Calculated values of $S$ (squares) and $S_-$ (triangles) versus the DFA exponent $\alpha_{DFA}$. The error bars indicate 
the region of one standard deviation ($\pm \sigma$). The horizontal line corresponds to $S_u$.}
\end{figure} 

\begin{figure}
\includegraphics{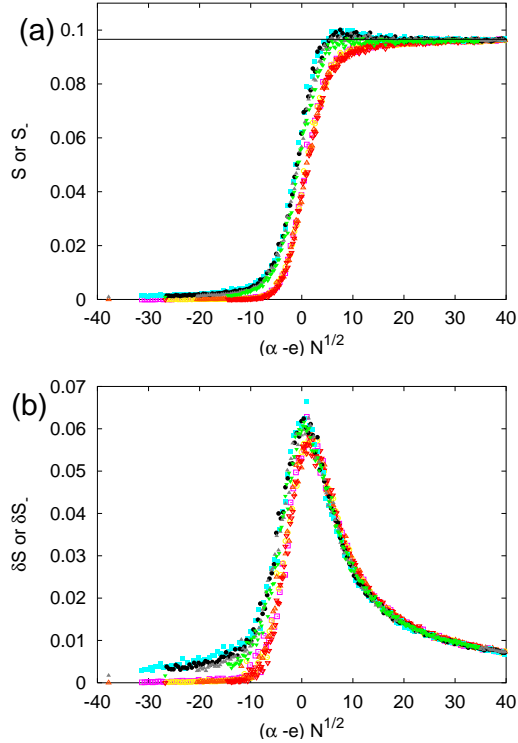}
\caption{\label{fig5} (Color Online) Calculated results for the on-off intermittency model discussed in the text: (a) depicts the average values of $S$ (closed symbols) and $S_-$ (open symbols) while (b) those of the fluctuations $\delta S$ and $\delta S_-$  versus the 
finite size scaling variable $(\alpha - \alpha_c) N^{1/2}$. The quantity $N$ stands for the number of the events considered in each sample time series; 
$N$=70000, 50000, 30000, 15000 correspond to squares, circles, triangles and inverted triangles, respectively. The horizontal line in (a) corresponds to $S_u$.}
\end{figure}

\begin{table}
\caption{The values of $S$ and $S_-$ together with the number of pulses $N$ for the SES activities (the original time series have lengths between $2 \times 10^3$ and $10^4$, compare Fig.1 with $f_{exp}$=1Hz) shown in Fig.1.}
\label{tab1}
\begin{ruledtabular}
\begin{tabular}{cccc}
Signal & $N$ & $S$ & $S_-$   \\
\hline
$M_1$ & $78\pm9$ & 0.094$\pm$0.005 & 0.078$\pm$0.003 \\
$M_2$ & $103\pm5$ & 0.089$\pm$0.003 & 0.084$\pm$0.003  \\
$M_3$ & $53\pm3$ & 0.089$\pm$0.004 & 0.093$\pm$0.004 \\
$M_4$ & $95\pm3$ & 0.080$\pm$0.005 & 0.086$\pm$0.006 \\
$V_1$ & $119\pm14$ & 0.078$\pm$0.006 & 0.092$\pm$0.005 \\
\end{tabular}
\end{ruledtabular}
\end{table}

\end{document}